# Reducing of phase retrieval errors in Fourier analysis of 2-dimensional digital model interferograms

**Jadranko Gladić, Zlatko Vučić, Davorin Lovrić**

Institute of Physics, Bijenička cesta 46, P. O. Box 304, 10001 Zagreb, Croatia

In order to measure the radial displacements of facets on surface of a growing spherical $Cu_{2-\delta}Se$ crystal with sub-nanometer resolution, we have investigated the reliability and accuracy of standard method of Fourier analysis of fringes obtained applying digital laser interferometry method. Guided by the realistic experimental parameters (density and orientation of fringes), starting from 2-dimensional model interferograms and using unconventional custom designed Gaussian filtering window and unwrapping procedure of the retrieved phase, we have demonstrated that for considerable portion of parameter space the non-negligible inherent phase retrieval error is present solely due to non-integer number of fringes within the digitally recorded image (using CCD camera). Our results indicate the range of experimentally adjustable parameters for which the generated error is acceptably small. We also introduce a modification of the (last part) of the usual phase retrieval algorithm which significantly reduces the error in the case of small fringe density.

## Introduction

Investigations by Ohachi's group[1,2,3] and our investigations[4,5,6] of superionic conductors, copper and silver chalcogenides, materials with high diffusivity of Cu and Ag atoms, enabled their promotion as new members of rather small group of materials that form equilibrium crystal shape (ECS), being the only ones that, besides $^4$He form ECS of macroscopic size (spherical faceted crystals of several mm in diameter). The size is crucial for investigating the non-equilibrium processes, i.e. shape kinetics during crystal growth as a function of the supersaturation forces driving the growth. Apart from thermodynamic reasons (approaching the limit of infinite size crystal), the size is specially important in the case of crystals growing at high temperatures at which the STM and AFM techniques are not applicable. Using two-beam optical interferometry (at growth velocities from 0.01 nm/s to few tens of nm/s, with declared resolution well below 10 nm) and by monitoring pressure of the liquid as a level gauge (0.05 to 50 μm/s) for investigating growth of $^4$He crystals at 2-250 mK, new facet growth modes were detected[7] the mechanism of which is not quite understood. On the other hand, our preliminary studies show that copper selenide ECS crystals at around 800 K exhibit similar facet growth modes. For their modeling the temperature complementarity of $^4$He and superionic conductors crystals seems to be of essential importance.

Our goal is to measure as precisely as possible the values of geometrical parameters of flat (facets) and curved parts of growing crystal surface necessary for characterizing the growth kinetics of our spherical apparently equilibrium single crystals in order to better understand the observed modes of growth, in particular the vertical and lateral facet growth. Using the flat and probably atomically smooth well developed facets on the surface of growing spherical crystal as highly reflecting objects, we apply digital laser interferometry method with refined[8] Fourier transform method of fringe pattern analysis to measure the displacements of facet perpendicular to the facet plane with sub-nanometer resolution.

In order to test the reliability and accuracy of the standard procedure of phase retrieval from the interference fringe pattern[7,8,9,10] by using the usual fringe Fourier analysis method, we have started[11] by testing the method itself applying it to an artificial, idealized fringe pattern in simple case where the 2-dimensional (2D) fringe pattern can be reduced to its 1-dimensional (1D) projection (where the $x$ axis is set to be perpendicular to the interferometric fringes). The method used as described in work of Kostianovski *et al.*[9] revealed the error in retrieved phase compared to the initially imposed phase in the model pattern. This inherent error was shown to originate from the non-integer number of fringes in the image field, and its dependence on spatial carrier frequency, initial phase setting and the deviation of number of fringes from integer value was investigated. We suggested a modification of (the last part of) the usual algorithm (applied after removal of carrier frequency from the fringe pattern Fourier spectrum and filtering it with an appropriately designed form of Gaussian filtering window) in order to reduce this error. The adapted algorithm was shown to have reduced the retrieved phase error more than threefold, specially for low wave vector values or small number of interferometric fringes within the observed image field.

Since under the usual experimental conditions it is extremely rare that the fringe pattern is obtained which can be treated within the 1D formalism already presented, we outline here the modifications necessary for treatment of more realistic general 2D case, showing the appropriate changes in design of the Gaussian filtering window and giving the explicit formulas for application in the real-life experimental situation which are shown to reduce the inherent phase determination errors. Their dependence on the parameters of experimentally obtained interferometric fringe patterns is clearly demonstrated.

## Fringe-pattern Fourier analysis

### *Interferometric Images*

During the crystal growth experiment as a rule we acquire a large number of 2D fringe patterns using a CCD camera. It takes pictures (frames) at regular time intervals (25 frames per second) of the part of growing crystal surface, including the facet with interference fringes, having generally elliptical shape (since the incident beam does not coincide with the normal to the facet). The background noise is automatically reduced by two orders of magnitude applying an original algorithm[8], the elliptical area with fringe pattern corresponding to the facet is detected and a square part of 64 x 64 pixels in size containing the interference fringes is taken from the facet area for further analysis. These interferograms (fringe patterns) are then used for extraction of the phase field value equivalent to changes in height of the corresponding facet area upon which the object interference beam is reflected.

A fringe pattern is, as described earlier[11] a set of straight parallel fringes with their intensity field $i(\vec{r})$ within 2D image modulated along wave vector $\vec{Q}$, the magnitude and direction of which are determined by the experimental setup (adjustable position of the reference mirror of the interferometer):

$$i(\vec{r}) = a(\vec{r}) + 2b(\vec{r})\cos\left[\vec{Q}\vec{r} + \varphi(\vec{r})\right], \qquad \vec{r} = x \cdot \vec{i} + y \cdot \vec{j} \qquad (1)$$

where $a(\vec{r})$ is the background intensity, $b(\vec{r})$ is the amplitude of modulation and $\varphi(\vec{r})$is the phase of the modulated signal.

Having in mind that we are actually dealing with a digital image from CCD camera, we stress that the field intensity $i(\vec{r})$ is not continuous, but is a set of $N$ x $M$ intensities separately recorded as pixels with 6-12 bit intensity resolution (depending on the CCD camera and frame grabber used).

In the general case of non-quadratic pixels, $x = nd_p^x$, $y = md_p^y$ where $d_p^x$ and $d_p^y$ are pixel dimensions, and $n$ and $m$ are number of pixels, in $x$ and $y$ direction, respectively. For $N$ x $M$ image field we have

$$i(x,y) = a(x,y) + 2b(x,y)\cos[(Q_x x + Q_y y + \varphi(x,y))] \qquad (2)$$

$$Q_x = \frac{2\pi}{Nd_p^x}(k_Q^x + D_Q^x),\quad Q_y = \frac{2\pi}{Md_p^y}(k_Q^y + D_Q^y),\qquad -\frac{N}{2} \le n \le \frac{N}{2} - 1,\qquad -\frac{M}{2} \le m \le \frac{M}{2} - 1$$

$$0 \le k_Q^x \le \frac{N}{2} - 1,\qquad -\frac{M}{2} \le k_Q^y \le \frac{M}{2} - 1,\qquad -0.5 < D_Q^x \le 0.5,\qquad -0.5 < D_Q^y \le 0.5$$

$$\varphi(x,y) = 2\pi\, \phi(x,y),$$

with the origin in the centre of the image field.

The spatial carrier frequency $\vec{Q}$ is expressed as the sum of an integer ($\vec{k}_Q$) and a non-integer ($\vec{D}_Q$) component in units of discrete Fourier space. The integers $k_Q^x$ and $k_Q^y$ are fixed during the entire experiment[8] (chosen by fine adjusting the experimental setup, measuring the density and orientation of fringes – the case for both $k_Q^x$ and $k_Q^y$ equal to zero has no physical significance), while the non-integer components oscillate in time from one image to the next.

We write again the Eq. (2) in the complex form (explicitly writing the dependence on discrete pixel coordinates $n$, $m$):

$$\begin{aligned} i(n,m) = a(n,m) + b(n,m)\exp\left\{ j2\pi\left[ \frac{(k_Q^x + D_Q^x)n}{N} + \frac{(k_Q^y + D_Q^y)m}{M} + \phi(n,m) \right]\right\} + \\ + b(n,m)\exp\left\{ - j2\pi\left[ \frac{(k_Q^x + D_Q^x)n}{N} + \frac{(k_Q^y + D_Q^y)m}{M} + \phi(n,m) \right]\right\} \end{aligned} \qquad (3)$$

## *Fourier Spectrum*

As suggested by Takeda *et al.*[10] the spatial frequency spectrum of the interferograms is obtained by Fourier transformation of Eq. (3) in 2 dimensions:

$$C(k^x,k^y) = FFT[i(n,m)] = \frac{1}{NM}\sum_{n=-N/2}^{N/2-1}\sum_{m=-M/2}^{M/2-1} i(n,m)\exp\left[-j2\pi(\frac{k^x}{N}+\frac{k^y}{M})\right] \quad (4)$$

$$-\frac{N}{2} \le k^x \le \frac{N}{2}-1, \qquad -\frac{M}{2} \le k^y \le \frac{M}{2}-1$$

(where $k^x$ and $k^y$ are integers), which results in the following expression:

$$\begin{aligned} C(k^x,k^y) = &\frac{1}{NM}\sum_{n=-N/2}^{N/2-1}\sum_{m=-M/2}^{M/2-1} b(n,m)\exp[-j2\pi\phi(n,m)]\exp\left\{-j2\pi\left[\frac{(k^x+k_Q^x+D_Q^x)n}{N}+\frac{(k^y+k_Q^y+D_Q^y)m}{M}\right]\right\}+ \\ &+\frac{1}{NM}\sum_{n=-N/2}^{N/2-1}\sum_{m=-M/2}^{M/2-1} a(n,m)\exp\left[-j2\pi\left(\frac{k^x n}{N}+\frac{k^y m}{M}\right)\right]+ \\ &+\frac{1}{NM}\sum_{n=-N/2}^{N/2-1}\sum_{m=-M/2}^{M/2-1} b(n,m)\exp[j2\pi\phi(n,m)]\exp\left\{-j2\pi\left[\frac{(k^x-k_Q^x-D_Q^x)n}{N}+\frac{(k^y-k_Q^y-D_Q^y)m}{M}\right]\right\} \end{aligned} \quad (5)$$

In order to proceed with an exact calculation we make an approximation by choosing that the magnitude of the background intensity $a$, the amplitude of the modulation $b$, and the phase $\phi$ are slowly varying quantities within the image frame and can be represented by their average values over the fringe pattern field: $a(n,m) = \bar{a}$, $b(n,m) = \bar{b}$, $\phi(n,m) = \bar{\phi}$. Then we have

$$
\begin{aligned}
C^{appr.}(k^x,k^y) &= \frac{\bar{b}}{NM}\exp\left(-j2\pi\bar{\phi}\right)\sum_{n=-N/2}^{N/2-1}\exp\left[-j\frac{2\pi}{N}(k^x+k_Q^x+D_Q^x)\,n\right]\sum_{m=-M/2}^{M/2-1}\exp\left[-j\frac{2\pi}{M}(k^y+k_Q^y+D_Q^y)\,m\right]+\\
&+\frac{\bar{a}}{NM}\sum_{n=-N/2}^{N/2-1}\exp\left(-j\frac{2\pi}{N}k^x n\right)\sum_{m=-M/2}^{M/2-1}\exp\left(-j\frac{2\pi}{M}k^y m\right)+\\
&+\frac{\bar{b}}{NM}\exp\left(j2\pi\bar{\phi}\right)\sum_{n=-N/2}^{N/2-1}\exp\left[-j\frac{2\pi}{N}(k^x-k_Q^x-D_Q^x)\,n\right]\sum_{m=-M/2}^{M/2-1}\exp\left[-j\frac{2\pi}{M}(k^y-k_Q^y-D_Q^y)\,m\right]
\end{aligned}
\tag{6}
$$

All three terms contain similar finite geometrical series of the type

$$
\sum_{l=-L/2}^{L/2-1}\exp\left(-j\frac{2\pi}{L}pl\right)=\exp\left(j\frac{\pi}{L}p\right)\mathrm{sinc}(p),\quad \mathrm{sinc}(p)=\frac{\sin(\pi\,p)}{\sin\left(\frac{\pi}{L}p\right)} \tag{7}
$$

for $L = N, M$ and $l = n, m$, respectively. The sinc functions are for $\vec{D}_Q \neq 0$ spread over the reciprocal space and their intensities are also significant in regions far from their corresponding maximum, even in the position of the respective maximum of the opposite sign (they are the source of intensity leakage to nearby spatial frequencies, just like in the 1D case – see Fig. 1 in our previous article[11]).

The final form of the Fourier spectrum is thus obtained:

$$
\begin{aligned}
C^{appr.}(k^x,k^y) &= \frac{\bar{b}}{NM}\exp\left(-j2\pi\bar{\phi}\right)\exp\left[j\pi\left(\frac{k^x+k_Q^x+D_Q^x}{N}+\frac{k^y+k_Q^y+D_Q^y}{M}\right)\right]\mathrm{sinc}\left(k^x+k_Q^x+D_Q^x\right)\mathrm{sinc}\left(k^y+k_Q^y+D_Q^y\right)+\\
&+\frac{\bar{a}}{NM}\exp\left[j\pi\left(\frac{k^x}{N}+\frac{k^y}{M}\right)\right]\mathrm{sinc}\left(k^x\right)\mathrm{sinc}\left(k^y\right)+\\
&+\frac{\bar{b}}{NM}\exp\left(j2\pi\bar{\phi}\right)\exp\left[j\pi\left(\frac{k^x-k_Q^x-D_Q^x}{N}+\frac{k^y-k_Q^y-D_Q^y}{M}\right)\right]\mathrm{sinc}\left(k^x-k_Q^x-D_Q^x\right)\mathrm{sinc}\left(k^y-k_Q^y-D_Q^y\right)
\end{aligned}
\tag{8}
$$

We again have three separate terms, corresponding to the first negative maximum, central maximum and the first positive maximum, respectively. The intensities in the Fourier space are strongly influenced by the behavior of the sinc functions (see Fig. 1 in Vučić, Gladić[11]), dependent on the non-integer part of the number of fringes contained within the field of pixels in both directions ($D_Q^x$, $D_Q^y$). When the number of fringes is exactly an integer ($\vec{D}_Q = 0$), there is no intensity leakage to spectral frequencies outside the well defined three maxima, and the spectrum is equal to zero everywhere except at the positions $\vec{k} = \pm\vec{k}_Q$ and $\vec{k} = 0$. For a non-integer number of fringes ($\vec{D}_Q \neq 0$) each of the terms representing the first (positive and negative) maxima is spread out over the entire Fourier space (visible "ridges" in Fig. 1). Thus the two complex conjugated first order maxima overlap at each point of the reciprocal space, the strength of overlapping being proportional to the size of $\vec{D}_Q$. Again, just like in 1D case[11], there is also significant intensity cutoff at the Nyquist frequency for $\vec{D}_Q \neq 0$.

## *Fourier Spectrum Shift and Removal of Carrier Frequency*

Following the usual procedure, in order to retrieve the phase of the interference signal, we now shift the Fourier spectrum in such a way that one of the side maxima is moved to the position $\vec{k} = 0$, thus removing the carrier frequency.

As we are dealing with discrete signal in the first place (intensities of pixels), the shift of the spectrum can be done exclusively by an integer number $\vec{k}_Q$. By formally replacing $k^x \to k^x + k_Q^x$, $k^y \to k^y + k_Q^y$, we obtain

$$\begin{aligned}
\overline{C}^{appr.}(k^x,k^y) = {} & \frac{\overline{b}}{NM}\exp\left[-j2\pi\left(\phi - \frac{D_Q^x}{2N} - \frac{D_Q^y}{2M}\right)\right]\exp\left[j\pi\left(\frac{k^x+2k_Q^x}{N} + \frac{k^y+2k_Q^y}{M}\right)\right]\mathrm{sinc}\left(k^x+2k_Q^x+D_Q^x\right)\mathrm{sinc}\left(k^y+2k_Q^y+D_Q^y\right)+ \\
& + \frac{\overline{a}}{NM}\exp\left[j\pi\left(\frac{k^x+k_Q^x}{N} + \frac{k^y+k_Q^y}{M}\right)\right]\mathrm{sinc}\left(k^x+k_Q^x\right)\mathrm{sinc}\left(k^y+k_Q^y\right)+ \\
& + \frac{\overline{b}}{NM}\exp\left[j2\pi\left(\phi - \frac{D_Q^x}{2N} - \frac{D_Q^y}{2M}\right)\right]\exp\left[j\pi\left(\frac{k^x}{N} + \frac{k^y}{M}\right)\right]\mathrm{sinc}\left(k^x - D_Q^x\right)\mathrm{sinc}\left(k^y - D_Q^y\right)
\end{aligned} \tag{9}$$

where $\overline{C}^{appr.}(k^x, k^y)$ denotes the shifted FFT spectrum of the 2D fringe pattern. The first of the three terms in this sum is the former negative first-order maximum, the second one is the former central maximum, and the third is the former positive first-order maximum, all shifted by the $\left(-k_Q^x, -k_Q^y\right)$ vector. The positive first-order maximum thus comes to the position near the origin, being left off-centre for the non-integral $\left(D_Q^x, D_Q^y\right)$.

## *Filtering using the 2-dimensional Gaussian Window*

In order to extract the phase information from the spectrum, the standard procedure is to isolate the shifted former first positive maximum from the rest of the spectrum. This is done by filtering of the Fourier spectrum by a narrow rectangular window centered at the origin ($\vec{k} = 0$) and apodized by a suitable Gaussian function, as suggested before[7,9,11]. The custom designed filtering function is centered at the origin of reciprocal space, again including the concept of the adjustable width depending on separation (positions) of first-order maxima relative to the origin and the edges of the Fourier space[11]. Its role is to reduce the intensity of the former central maximum (now shifted to $-\vec{k}_Q$) as well as the former first negative maximum (now shifted to $-2\vec{k}_Q$). It is also, in analogy with the 1D case used to remove the intensity jumps formerly located at the Nyquist frequencies (the edges of the field now at $N/2-1-\left|k_Q^x\right|$ and $N/2-1-\left|k_Q^y\right|$).

For $N = M$ (the usual experimental case of quadratic sampled intensity field of $N$ x $N$ pixels), the Gaussian window of the following form fulfills all the mentioned requirements:

$$GW_{2D} = \exp\left\{-\alpha \ln 10 \frac{\left(\frac{N}{2}-2\right)^2\left(\frac{N}{2}-1\right)^2}{\left[\left(k_Q^x\right)^2+\left(k_Q^y\right)^2\right]\left(\frac{N}{2}-1-\left|k_Q^x\right|\right)^2\left(\frac{N}{2}-1-\left|k_Q^y\right|\right)^2} k^2\right\} \tag{10}$$

The factor $\alpha$ determines the level to which the former central maximum is reduced in intensity, i.e. it is multiplied by a factor $10^{-\alpha}$ ($\alpha \geq 3$) and suppressed at least to a level comparable with noise. (The results weakly depend on the value of $\alpha$.)

### *Inverse Fast Fourier Transformation and Phase Extraction*

Denoting $\psi = \phi - D_Q^x / 2N - D_Q^y / 2M$ we now have

$$\overline{C}_W^{appr.}(k^x, k^y) = \frac{\overline{b}}{NM} \left\{ \begin{array}{l} \exp(-j2\pi\psi)\exp\left[ j\pi\left( \frac{k^x + 2k_Q^x}{N} + \frac{k^y + 2k_Q^y}{M} \right) \right] \mathrm{sinc}\left(k^x + 2k_Q^x + D_Q^x\right)\mathrm{sinc}\left(k^y + 2k_Q^y + D_Q^y\right)\, GW_{2D} + \\ \exp(j2\pi\psi)\exp\left[ j\pi\left( \frac{k^x}{N} + \frac{k^y}{M} \right) \right] \mathrm{sinc}\left(k^x - D_Q^x\right)\mathrm{sinc}\left(k^y - D_Q^y\right)\, GW_{2D} \end{array} \right\} + \\ + \mathrm{occ(FCM)} \tag{11}$$

The former central maximum is after applying the filter function (Gaussian window) reduced in amplitude over 1000 times, and is therefore denoted as occ(FCM) and neglected in further calculations.

In the real experiment, by adjusting the optical elements we choose the density and orientation of fringes, thus we select $\left(k_Q^x, k_Q^y\right)$ which remains constant[8] during the observed crystal growth, so that the filtering function is indeed well defined.

The influence of the leakage effect (Fig. 1, also Fig. 1 in ref. [11]) is still clearly visible through the presence of sinc functions in Eq. (11), originating, as already stressed, as a direct consequence of $\vec{D}_Q \neq 0$, which cannot be removed by the application of Gaussian filter, no matter how strong and narrow it would be. Differing from the analysis of Kostianovski *et al.*[3], we see that besides the former first positive maximum contribution which is usually expected to be sufficient for extracting the phase information, there is still a contribution in the spectrum coming from the former first negative maximum as well, which was completely neglected in previous work[9].

Following the standard recipe, just like in the 1D case[11], the next step would be the application of the inverse Fourier transform to the shifted and filtered spectrum $\overline{C}_W^{appr.}(k^x,k^y)$, to obtain a complex function $I(n,m)$ in the real space from which the phase would be extracted. The following ubiquitous formula for extracting the phase field from the inverse Fourier transform is not accurate enough:

$$\phi = \frac{D_Q^x}{2N} + \frac{D_Q^y}{2M} + \frac{1}{2\pi}\arctan\left\{\frac{\mathrm{Im}[I(n,m)]}{\mathrm{Re}[I(n,m)]}\right\} \qquad (12)$$

$$I(n,m) = \mathrm{IFFT}\left[\overline{C}_W^{appr.}(k^x,k^y)\right] = \frac{1}{NM}\sum_{k^x=-N/2}^{N/2-1}\sum_{k^y=-M/2}^{M/2-1}\overline{C}_W^{appr.}(k^x,k^y)\exp\left[j2\pi\left(\frac{k^x n}{N}+\frac{k^y m}{M}\right)\right] \qquad (13)$$

There is also the 2D analogue of the linear phase correction discussed in our previous work[8,11], $D_Q^x/2N + D_Q^y/2M$, independent of carrier frequency, inversely proportional to the dimensions of the image field *N* x *M*.

## Results and discussion

As discussed for the case in which the phase field was reduced to 1D representation[11], the main disadvantage of the method is the *x-y* resolution problem originating from the artificial retrieved phase field modulation introduced inevitably from the incomplete shift (by the integer valued vector) of the first positive maximum of the Fourier spectrum towards the origin, thus being left $\vec{D}_Q$ off center. In order to determine the retrieved phase error as compared with the initially introduced phase, we are once again forced to do the averaging over the entire retrieved phase field.

In the real experimental data the non-integer $\vec{D}_Q$ changes from one image to the next[8], so that its influences on the retrieved phase field modulation do not cancel out when calculating the changes in the phase field between consecutive images. Therefore, it is not possible to obtain the reconstruction of fine details of the growing crystal facet – the only reliable information one can get is about the change of position of the growing facet calculated from averaged phase field over the entire *N* x *M* (typically 64 x 64 pixels) selected fringe pattern area. Taking these regions from different parts of the facet image can give some local information about the surface profile changes, the *x*-*y* resolution being given by the size of the crystal surface corresponding to the selected region (depending on the magnification of the experimental setup).

Before averaging, because of these same artificial modulations, the appropriate 2D unwrapping must be applied. Fortunately, the simple methods found in standard mathematical software prove to be quite appropriate (but should be applied with care, consecutively first to columns, then to rows of the matrix), except in few rare occasions where the previous application of an "initial preparatory" unwrapping following the algorithm by Abbas[12] helps to remove some problematic phase jumps.

The difference of thus retrieved phase (by evaluating Eq. (12), averaged over the phase field, taking the maximum possible non-integer part of the spatial carrier frequency $\vec{Q}$ ($D_Q^x = 0.5$ and $D_Q^y = 0.5$)) and the initially given phase $\phi_{in}$ (in Eq. (1)) is equal to the one shown in Fig. 2. of our previous article[11] (for *N* = *M* = 64, $\bar{b}/\bar{a} = 0.5$, α = 3 and $k_Q^y = 0$), as it should be. The dependence of the error magnitude on initially set phase and on the integer part of the spatial carrier frequency $\vec{k}_Q$ for these parameters shows maximum value for the input phase $\phi_{in}$ = ±0.155, being equivalent to 10 nm frontal displacement ($\Delta\bar{z}$) of the observed growing facet (λ = 632.8 nm for He-Ne laser, the angle between the incident and the reflected beam θ = $0^0$, using the formula

$$\Delta\bar{z} = \Delta\bar{\phi}\,\frac{\lambda}{4\pi\cos\frac{\theta}{2}} \qquad (14)$$

where $\Delta\bar{\phi}$ is generally the difference between values of the averaged unwrapped retrieved phase field from two consecutive interferograms).

The dependence of the retrieved phase error (using standard Eq. (12)) on integer part components of the spatial carrier frequency $k_Q^x$ and $k_Q^y$ is shown in Fig. 2 (for the input phase $\phi_{in}$ = 0.155 that produces maximum error, and for the maximum possible departure from the integer values $D_Q^x = 0.5$ and $D_Q^y = 0.5$).

Dependence of the averaged retrieved phase error value (using standard Eq. (12)) on non-integer part of the spatial carrier frequency $\vec{D}_Q = \left(D_Q^x, D_Q^y\right)$ is shown by larger circles in Fig. 3., within a small part of the Fourier space. The error vanishes for $\vec{D}_Q = 0$, and is showing up with departure from the integer value of the carrier frequency, diminishing as the $k_Q^y$ component grows.

## Corrected phase extraction

By introducing the $\overline{C}_W^{appr.}(k^x, k^y)$ from Eq. (11) into the expression (13) for $I(n,m)$ and omitting the occ(FCM) term, we have:

$$I(n,m) = \frac{\bar{b}}{NM}\left\{\begin{array}{l} \exp(-j2\pi\psi)\sum_{k^x=N/2}^{N/2-1}\exp\left(j2\pi\frac{k^x n}{N}\right)\exp\left(j\pi\frac{k^x+2k_Q^x}{N}\right)\mathrm{sinc}\left(k^x+2k_Q^x+D_Q^x\right)\sum_{k^y=M/2}^{M/2-1}\exp\left(j2\pi\frac{k^y m}{M}\right)\exp\left(j\pi\frac{k^y+2k_Q^y}{M}\right)\mathrm{sinc}\left(k^y+2k_Q^y+D_Q^y\right)GW_{2D}+ \\ \exp(j2\pi\psi)\sum_{k^x=N/2}^{N/2-1}\exp\left(j2\pi\frac{k^x n}{N}\right)\exp\left(j\pi\frac{k^x}{N}\right)\mathrm{sinc}\left(k^x-D_Q^x\right)\sum_{k^y=M/2}^{M/2-1}\exp\left(j2\pi\frac{k^y}{M}\right)\exp\left(j\pi\frac{k^y}{M}\right)\mathrm{sinc}\left(k^y-D_Q^y\right)GW_{2D} \end{array}\right\}$$

(15)

Sums in both terms are the inverse Fourier transforms of mutually similar functions *K* and *O*, hereby explicitly listed by components ($i = x, y$):

$$K^i(k^i) = \exp\left( j\pi \frac{k^i + 2k_Q^i}{N} \right) \mathrm{sinc}\left(k^i + 2k_Q^i + D_Q^i\right) GW_{2D}^i ,$$

$$O^i(k^i) = \exp\left( j\pi \frac{k^i}{N} \right) \mathrm{sinc}(k^i - D_Q^i) GW_{2D}^i \qquad (16)$$

Using this notation we have

$$I(n,m) = \frac{\bar{b}}{N^2 M^2} \begin{Bmatrix} \exp(-j2\pi\psi)\,\mathrm{IFFT}\,K^x(n)\,\mathrm{IFFT}\,K^y(m) + \\ \exp(j2\pi\psi)\,\mathrm{IFFT}\,O^x(n)\,\mathrm{IFFT}\,O^y(m) \end{Bmatrix} \qquad (17)$$

It can be shown that $K = K^x K^y$ and $O = O^x O^y$ ($GW_{2D} = GW_{2D}^x GW_{2D}^y$). Solving this equation as a linear system of equations for the sine and the cosine of the corrected phase $\psi = \phi - D_Q^x / 2N - D_Q^y / 2M$, by using the real and imaginary parts of these functions, we finally obtain a somewhat complicated, but more accurate expression for phase retrieval

$$\tan 2\pi\psi = \frac{\mathrm{Re}[\mathrm{IFFT}(O+K)]\,\mathrm{Im}[\mathrm{IFFT}(I)] - \mathrm{Im}[\mathrm{IFFT}(O+K)]\,\mathrm{Re}[\mathrm{IFFT}(I)]}{\mathrm{Re}[\mathrm{IFFT}(O-K)]\,\mathrm{Re}[\mathrm{IFFT}(I)] + \mathrm{Im}[\mathrm{IFFT}(O-K)]\,\mathrm{Im}[\mathrm{IFFT}(I)]} \qquad (18)$$

$$\phi = \frac{D_Q^x}{2N} + \frac{D_Q^y}{2M} + \frac{1}{2\pi} \arctan\left\{ \frac{\mathrm{Re}[\mathrm{IFFT}(O+K)]\,\mathrm{Im}[\mathrm{IFFT}(I)] - \mathrm{Im}[\mathrm{IFFT}(O+K)]\,\mathrm{Re}[\mathrm{IFFT}(I)]}{\mathrm{Re}[\mathrm{IFFT}(O-K)]\,\mathrm{Re}[\mathrm{IFFT}(I)] + \mathrm{Im}[\mathrm{IFFT}(O-K)]\,\mathrm{Im}[\mathrm{IFFT}(I)]} \right\} \qquad (19)$$

which is completely analogous to the previously given[11] expression for the 1D case. This introduces some more computational effort as compared to standard (less accurate expression (12)), but the improvement is obvious from the Fig. 3, where the phase values obtained from this expressions are shown with smaller grey circles. The error is still existing, and is dependent on non-integer carrier frequency (no error for $\vec{D}_Q = 0$), but its amplitude is significantly reduced. It can be further suppressed by increasing the value of $\alpha$ in the Gaussian window definition (taking care that it still makes sense, concerning the noise present in realistic experiments – the removal of noise from original data[8] is an obligatory prerequisite for high quality measurements).

In analogy with Fig. 2, the dependence of the retrieved phase error using this corrected expression given by Eq. (19) on integer part components of the spatial carrier frequency $k_Q^x$ and $k_Q^y$ is shown in Fig. 4 (again for the worst possible case of $\phi_{in}$ = 0.155, and for the maximum values $D_Q^x = 0.5$ and $D_Q^y = 0.5$).

In order to emphasize the significance of the improved (corrected) expression for phase retrieval and to demonstrate its meaning in real experimental situations, we show in Fig. 5. two cross-sections of combined Figs. 2. and 4. In the upper part we show the behavior of retrieved phase error in the case of interferometric fringes exactly parallel to the image *x*-axis ($k_Q^y = 0$), while in the lower part we show the error in the case of fringes running diagonally over the image field ($k_Q^x = k_Q^y$). The error of the retrieved phase (averaged over the whole field) is shown in nanometers – it is in fact the error in determination of the displacement of the growing facet in the direction perpendicular to its plane (for λ = 632.8 nm for He-Ne laser, the angle between the incident and the reflected beam of $0^0$). Diamonds show the error obtained using the standard Eq. (12), and squares the corrected value according to Eq. (19). The size of the errors is clearly suppressed several times by using the corrected expression, specially for the small density of fringes within the image field. The error itself is obviously smaller for fringes running diagonally, and is almost negligible in any case for not too small and not too big density of fringes. (The insets show the enlarged view of the interval of $k_Q^x$ values between 6 and 26, where the error is smaller than ±0.35 nm for horizontal fringes, and another order of magnitude smaller for diagonal fringes, $k_Q^x = k_Q^y$. In our case, growth of a single atom layer on the (111) facet of $Cu_{2-x}Se$ is equal to displacement of 0.35 nm.) As these parameters (number and orientation of fringes) can be adjusted by fine tuning the elements of the experimental setup, these conclusions can be considered as guidelines for performing more reliable and accurate measurements.

## Conclusions

Striving to measure as precisely as possible the values of radial displacements of facets on the surface of growing spherical single crystals of $Cu_{2-\delta}Se$ using digital laser interferometry method, we have undertaken to test the reliability and accuracy of standard Fourier analysis procedure for phase retrieval from interference fringe patterns. Building upon our previous work on idealized fringe pattern which can be reduced to its 1D projection[11], we extended our study to more realistic, quite general 2D case of digitally recorded interferograms, investigating wide range of experimentally adjustable parameters (fringe density and orientation) on a model interferogram. We demonstrated the existence of inherent errors in the retrieved phase due exclusively to the non-integer number of interferometric fringes within the observed digitally sampled image obtained from a CCD camera. Within the framework of the usual procedure (Fourier fringe analysis) for retrieving the phase value, we introduced fully adjustable custom designed 2D Gaussian filtering window. The appropriate 2D unwrapping of the retrieved phase field is applied as consecutive multiple-stage unwrapping algorithm. The obtained phase field must be averaged over the entire selected interferogram area. Comparing the retrieved phase with the initially set one, we showed the dependence of the retrieved phase errors on the input phase value, fringe density and orientation (for generally non-integer 2D spatial carrier frequencies). We presented a modified expression to replace the one usually used for phase retrieval, which results in significant reduction of the error amplitude, particularly for small fringe density values. Our results show that there is a range of experimentally selectable parameters within which the retrieved phase error, expressed as corresponding observed facet displacement, is smaller than ±0.35 nm in the case of interferometric fringes running parallel to *x*-axis of the interferogram, and even smaller than ±0.035 nm in the case of diagonally oriented fringes.

We acknowledge gratefully the financial support of the Ministry of Science, Education and Sport of the Republic of Croatia.

## References

1. T. Ohachi, I. Taniguchi, "Growth of α-$Ag_2S$ and α-$Ag_2Se$ single crystal through a capillary tube", in *Fast Ion Transport in Solids, Electrodes and Electrolytes*, P. Vashishta, J. N. Mundy and G. K. Shenoy, eds.(North Holland, New York, Amsterdam, Oxford), pp. 597-600.
2. T. Ohachi, I. Taniguchi, "Roughening transition for the ionic-electronic mixed superionic conductor α-$Ag_2S$", J. Cryst. Growth **65** (1983) 84-88.
3. T. Ohachi, S. Imai, T. Tanaka, H. Yamai, I. Taniguchi, "Semiconducting and atomic properties of the mixed conductor α-$Ag_2S$", Solid State Ionics **28-30** (1988) 1160-1166.
4. Z. Vučić, and J. Gladić, "Shape relaxation during equilibrium - like growth of spherical cuprous selenide single crystals", Fizika A (Zagreb) 9(1), 9 – 26 (2000) (http://fizika.phy.hr/fizika_a/av00/a9p009.htm).
5. Z. Vučić, and J. Gladić, "Growth rate of equilibrium - like – shaped single crystals of superionic conductor cuprous selenide", J. Crystal Growth 205, 136 – 152 (1999).
6. J. Gladić, Z. Vučić, and D. Lovrić, "Critical behavior of the curved region near 111 – facet edge of equilibrium shape cuprous selenide large single crystals", J. Crystal Growth 242, 517 – 532 (2002).
7. J. P. Ruutu, P. J. Hakonen, A. V. Babkin, A. Zu. Parshin and G. Tvalashvili, "Growth of 4He - crystals at mK – temperatures", J. Low Temp. Phys. 112, 117-164 (1998).
8. D. Lovrić, Z. Vučić, J. Gladić, N. Demoli, S. Mitrović, and M. Milas, "Refined Fourier – transform method of analysis of full 2D digitized interferograms", Appl. Opt. 42(8), 1477 – 1484 (2003).
9. S. Kostianovski, S. G. Lipson, and E. N. Ribak, "Interference microscopy and Fourier fringe analysis applied to measuring spatial refractive – index distribution", Appl. Opt. 32(25), 4744 – 4755 (1993).

10. M. Takeda, H. Ina, and S. Kobayashi, "Fourier – transform method of fringe – pattern analysis for computer – based topography and interferometry", J. Opt. Soc. Am. 72(1), 156-160 (1982).

11. Z. Vučić, J. Gladić, "Phase retrieval errors in standard Fourier fringe analysis of digitally sampled model interferograms", Appl. Opt. 44, 6940-6947 (2005)

12. Kattoush Abbas, "A New Recurrent Approach for Phase Unwrapping", International Journal of Applied Science and Engineering 3(2), 135-143 (2005)

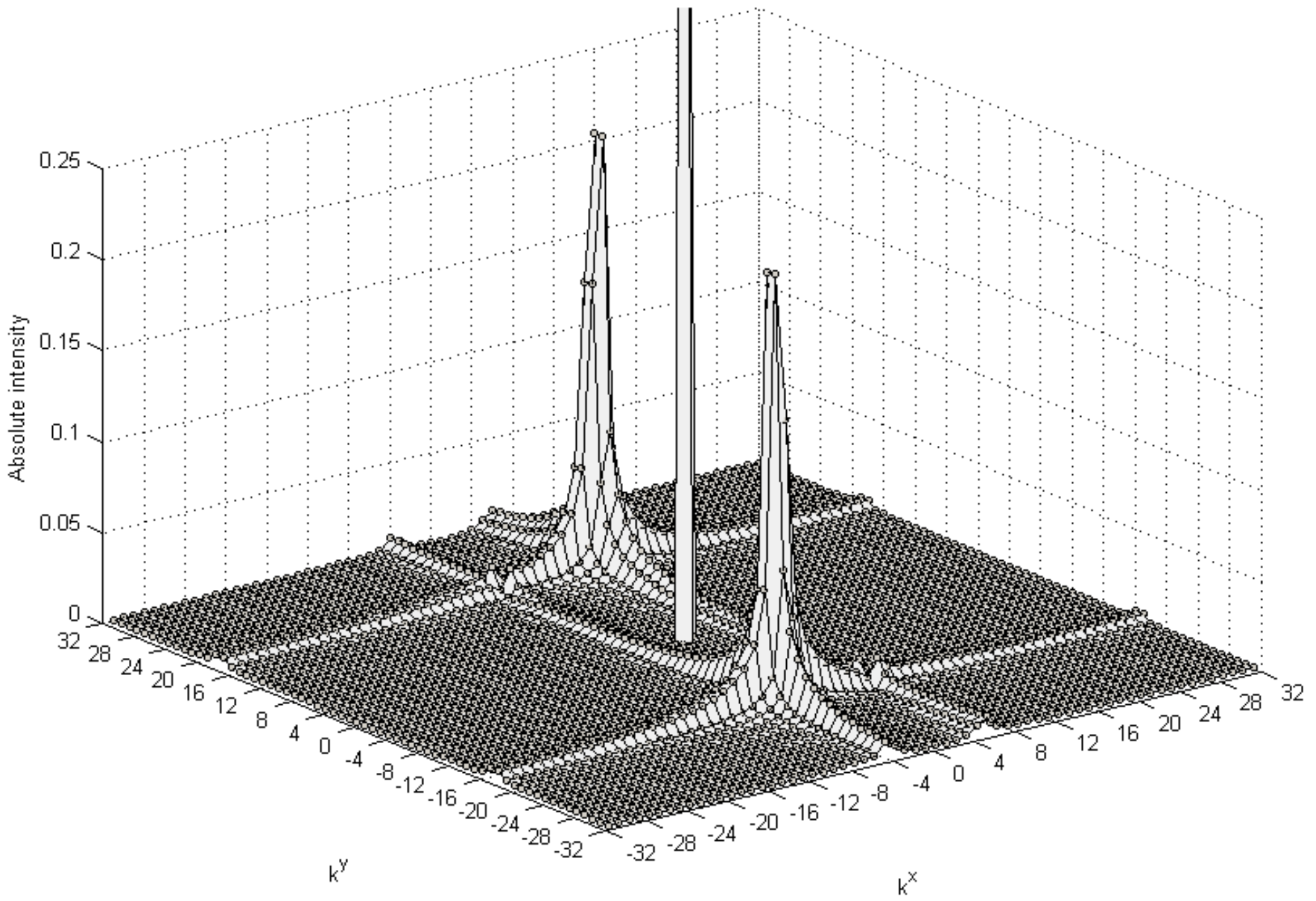

Fig. 1. The leakage effect originating from the sinc functions in Eq. (8) in the Fourier spectrum with shown narrow central maximum and positive and negative first order maxima (enlarged). Parameters are: $N = M = 64$, $\bar{b}/\bar{a} = 0.5$, $\phi = 0.155$, $k_Q^x = 5$, $k_Q^y = 17$, $D_Q^x = 0.5$ and $D_Q^y = 0.5$.

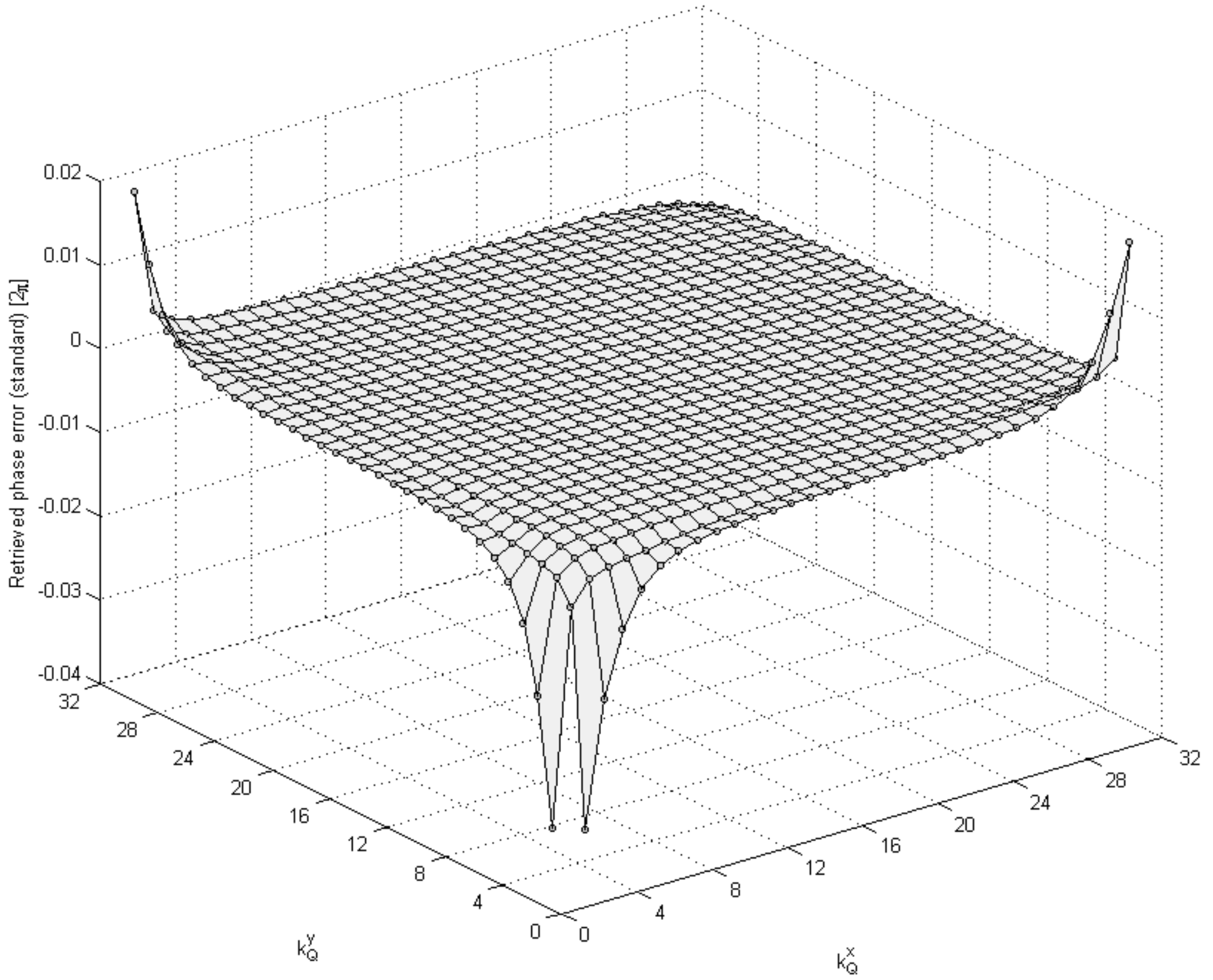

Fig. 2. Difference between the retrieved phase (from Eq. (12)) and the initially set phase, in units of 2π, as a function of integer spatial frequency components $k_Q^x$ and $k_Q^y$. Parameters are: $N = M = 64$, $\overline{b}/\overline{a} = 0.5$, $\phi_{in} = 0.155$, α = 3, $D_Q^x = 0.5$ and $D_Q^y = 0.5$.

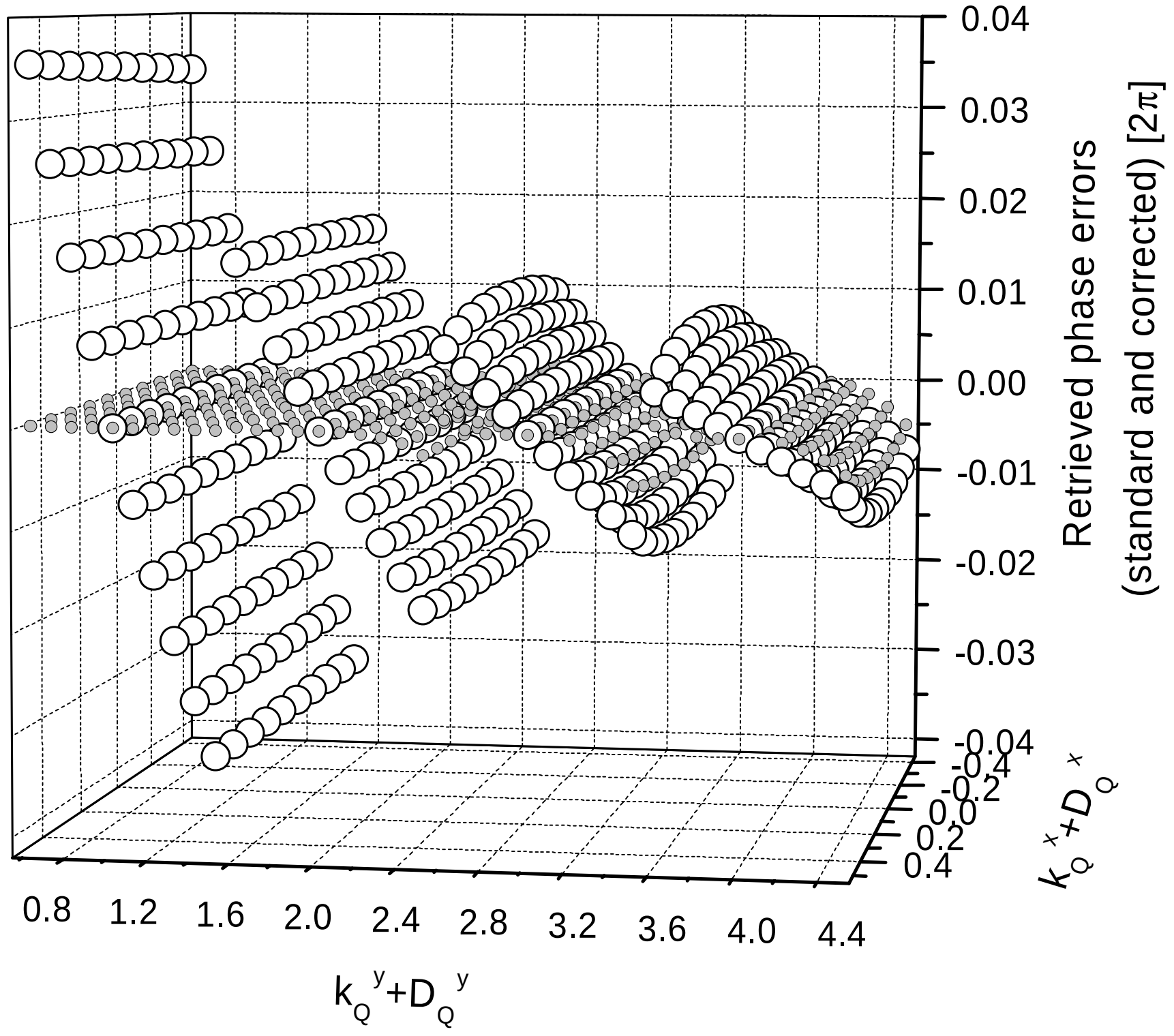

Fig. 3. Difference between the retrieved phase (from Eq. (12)) and the initially set phase, in units of $2\pi$, as a function of integer and non-integer parts of spatial carrier frequency components, shown in a small part of Fourier space. Larger white circles show the retrieved phase error obtained from standard Eq. (12), and the small grey circles show the corrected retrieved phase error obtained from Eq. (19). Parameters are: $N$ = $M$ = 64, $\bar{b}/\bar{a}$ = 0.5, $\phi_{in} = 0.155$, α = 3.

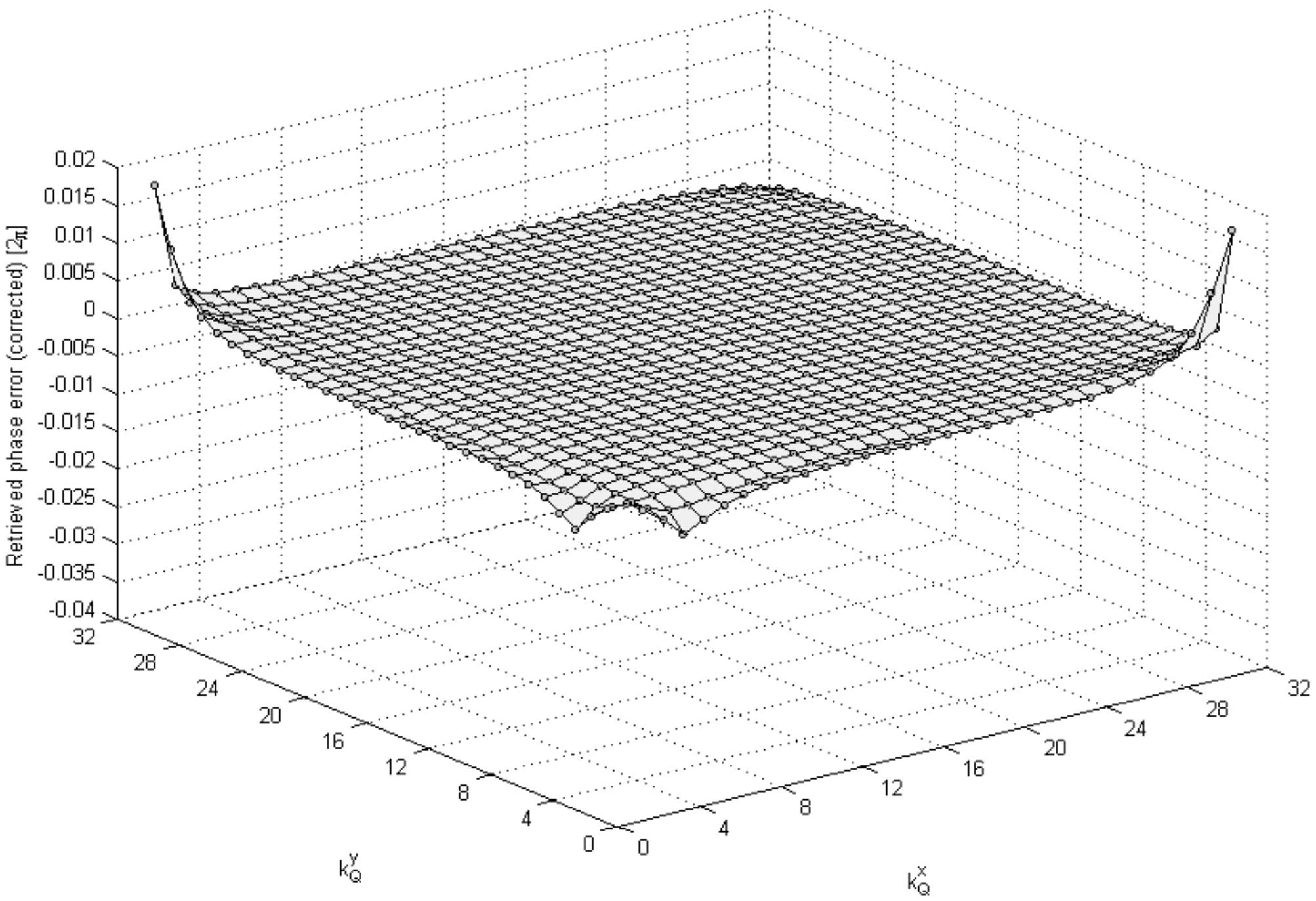

Fig. 4. Difference between the retrieved phase (from Eq. (19)) and the initially set phase, in units of 2$\pi$, as a function of integer spatial frequency components $k_Q^x$ and $k_Q^y$ (compare with Fig. 2.). Parameters are: $N = M = 64$, $\bar{b}/\bar{a}$ = 0.5, $\phi_{in} = 0.155$, α = 3, $D_Q^x$ = 0.5 and $D_Q^y$ = 0.5.

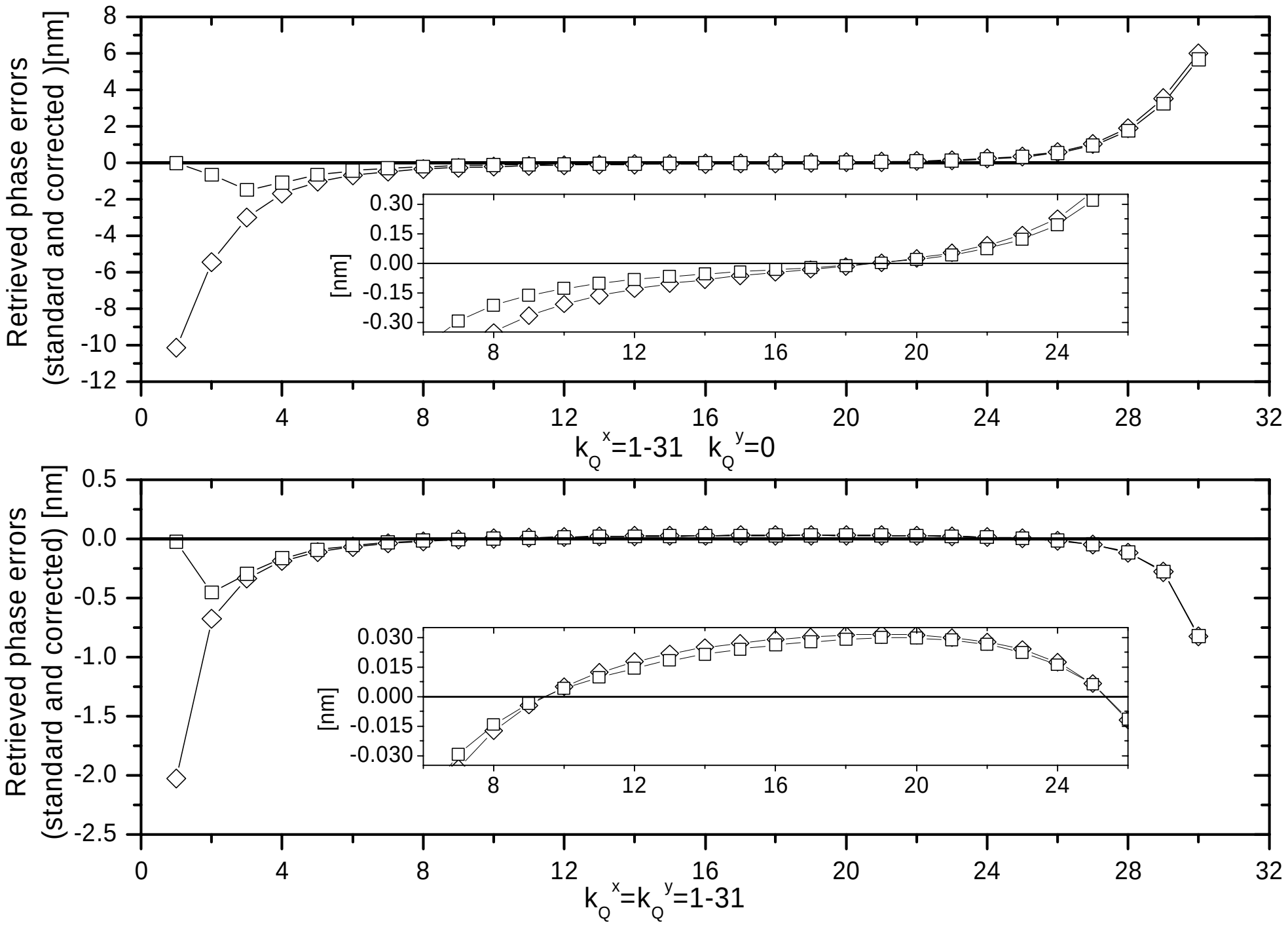

Fig. 5. Comparison of the retrieved phase errors (given in nm as the corresponding displacement of the observed object) using standard ◊ (Eq. (12)) and corrected □ (Eq. (19)) expressions. The upper panel is for interferometric fringes parallel to the *x*-axis, and the lower panel for fringes running diagonally. The insets show enlarged view of parameter range within which the errors are practically negligible. Parameters are: $N = M = 64$, $\bar{b}/\bar{a} = 0.5$, $\phi_{in} = 0.155$, $\alpha = 3$, $D_Q^x = 0.5$ and $D_Q^y = 0.5$.